\documentstyle[preprint,aps,psfig]{revtex}
\newcommand{\T}{{\large$t$}}

\begin{document}
\draft
\preprint{MKPH-T-96-32}
\title{A Dispersion Theoretical Approach to the Threshold Amplitudes 
of Pion Photoproduction}

\author{O.~Hanstein, D.~Drechsel and L.~Tiator}
\address{Institut f\"ur Kernphysik, Universit\"at Mainz, 55099 Mainz, Germany}
\date{\today}
\maketitle

\begin{abstract}
  We give predictions for the partial wave amplitudes of pion
  photoproduction near threshold by means of dispersion relations at
  fixed $t$. The free parameters of this approach are determined by
  a fit to experimental data in the energy range 160 MeV$ \le
  E_{\gamma} \le $420 MeV. The observables near threshold are
  found to be rather sensitive to the amplitudes in the resonance
  region, in particular to the $\Delta$ (1232) and $N^*$ (1440). We
  obtain a good agreement with the existing threshold data for both
  charged and neutral pion production. Our predictions also agree
  well with the results of chiral perturbation theory, except for
  neutral pion production off the neutron.

\end{abstract}

\section{Introduction} \label{sec:intro}

The photoproduction of pions near threshold has been a topic of
considerable experimental and theoretical activities over the past
years, ever since the Saclay \cite{Maz86} and Mainz \cite{Bec90} groups 
showed that
the data for the reaction $\gamma p \rightarrow p \pi^0$ were at variance
with the predictions of a low energy theorem (LET), which 
was derived in the early 70's
by de Baenst \cite{deB70}
and Vainshtein and Zakharov \cite{Vai72}. Being based on fundamental principles
like Lorentz, gauge and chiral
symmetry, this LET predicted the threshold amplitudes in form of a power series
in $\mu = m_{\pi}/m_N$, with $m_{\pi}$ and $m_N$ the masses of pion and 
nucleon, respectively. The discrepancy between the theoretical expectations
and the experimental data was eventually explained by 
Bernard et al. \cite{Ber91}
by an explicit one-loop calculation in (relativistic) ChPT. It was shown that 
the flaw of the LET was in the assumption that the amplitudes 
would be analytical
functions in $\mu$. In fact, they contain logarithmic terms which cancel in
the final result but whose expansion leads to previously unexpected 
contributions to the power series in $\mu$. In the following years, 
calculations were considerably refined by introducing the heavy baryon
formalism to ChPT, leading to a well-defined expansion in $\mu$, 
and by including higher orders in the chiral expansion, up to order
$p^4$ in the $s$-wave multipole $E_{0+}$ and $p^3$ in the $p$-wave
multipoles \cite{Ber96a}. To that order there appear 3 low-energy 
constants, which 
have to be fitted to the data or estimated by resonance saturation. While 
there is some doubt concerning the convergence of the loop expansion
for the $s$-waves, the expansion for the $p$-waves seems to converge
rather well. Moreover, two combinations of the $p$-waves are free of
low energy constants, leading to new LETs to order $p^3$.
On the experimental side, new precision data have been
obtained by both the TAPS collaboration at Mainz \cite{Fuc96}
and the Sascatoon \cite{Ber96} group.
These experiments come closer to threshold than previously possible and
clearly show that the strength of the $s$-wave multipole is reduced by nearly
a factor 2 as compared to the old LET. 
The calculations of ChPT describe the data quite nicely. However, there 
remains the question of the convergence of the loop expansion. 

Concerning the production of  charged pions, the theoretical description is in
much better shape. The dominant contribution near threshold is the
Kroll-Ruderman term \cite{Kro54}, and corrections up to $O(\mu^3)$ have 
recently been 
calculated in ChPT \cite{Ber96b}. In that case loop corrections are fairly 
small,
which offers the possibility to determine the pion-nucleon coupling constant 
$g_{\pi N}$ by new precision experiments.

It is the aim of this letter to investigate the threshold production of 
neutral and charged pion by means of dispersion relations at fixed $t$. 
The free
parameters of this approach have been fitted to experiments in the energy 
range from 160 MeV to about 420 MeV, i.e. we predict the threshold 
behaviour on the basis of the data at the higher energies.
The paper is organized as follows. In section~\ref{sec:thr}, 
we summarize the kinematics
and the observables for pion photoproduction. Some information on dispersion
relations at fixed $t$ is given in section~\ref{sec:disp}, together 
with a brief discussion
of the assumptions and parameters involved in our calculations. We then discuss
the predictions of dispersion theory for threshold pion production in 
section~\ref{sec:results}.
Finally, we summarize our findings and present some conclusions
in section~\ref{sec:concl}.

\section{Threshold pion photoproduction} \label{sec:thr}

We consider the reaction
\begin{equation}
\gamma (q) + N (p) \rightarrow \pi (k) + N' (p').
\end{equation}
The 3-momenta of photon and pion in the $cm$ frame will be denoted by 
$\vec{q}$ and $\vec{k}$, respectively. We will use $\Theta_{cm} = 
\Theta$ to describe the scattering angle, $W$ for the $cm$ energy of the
$\pi N$ system, and $E_{\gamma}$ for the lab energy of the incident 
photon. In the following we will give the multipole expansion for the 
pertinent observables taking into account only the $s$-wave multipoles 
$E_{0+}$ and the 3 $p$-waves $E_{1+}$, $M_{1+}$ and $M_{1-}$. 
We note, however, that our numerical results in section \ref{sec:results} 
include the
higher partial waves as well. For convenience we also introduce the 
combinations
\begin{equation} \label{P1_P3}
P_{1} = 3E_{1+} + M_{1+} - M_{1-}, \hspace{3mm}
P_{2} = 3E_{1+} - M_{1+} + M_{1-}, \hspace{3mm}
P_{3} = 2M_{1+}+M_{1-}.
\end{equation}

With these definitions and approximations, the differential cross section
in the $cm$ frame reads \cite{Dre92}
\begin{equation} \label{dsig}
\frac{d\sigma}{d\Omega} = \frac{\mid\vec k\mid}{\mid\vec q\mid}
(A + B \cos\theta + C \cos^{2}\theta),
\end{equation}

where

\begin{eqnarray} \label{ABC}
A & = & \mid E_{0+}\mid^{2} + \frac{1}{2}\mid P_{2}\mid^{2} 
      + \frac{1}{2}\mid P_{3}\mid^{2}, \nonumber \\
B & = & 2{\mathrm Re} (E_{0+}P_{1}^{\ast}), \\
C & = & \mid P_{1}\mid^{2} - \frac{1}{2}\mid P_{2}\mid^{2}
      - \frac{1}{2}\mid P_{3}\mid^{2}, \nonumber
\end{eqnarray}

leading to the total cross section

\begin{equation} \label{sig_tot}
\sigma_{\mathrm tot}  =
 4\pi\frac{\mid\vec k\mid}{\mid\vec q\mid}\left(\mid E_{0+}\mid^{2}
 + \mid M_{1-}\mid^{2} + 6\mid E_{1+}\mid^{2} + 2\mid M_{1+}\mid^{2}\right).
\end{equation}

The single-polarization observables are the beam asymmetry,

\begin{equation}
  \label{Sigma}
  \Sigma = \Gamma \sin\theta(\mid P_{3}\mid^{2}-\mid P_{2}\mid^{2}),
\end{equation}

the target asymmetry

\begin{equation}
  \label{targ_asym}
  T = 2\Gamma \hspace{1mm}{\mathrm Im}
      \left((E_{0+}+\cos\theta P_{1})(P_{3}-P_{2})^{\ast}\right),
\end{equation}

and the recoil polarization

\begin{equation}
  \label{rec_pol}
  P = 2\Gamma \hspace{1mm}{\mathrm Im}
       \left((E_{0+}+\cos\theta P_{1})(P_{3}+P_{2})^{\ast}\right),
\end{equation}

where

\begin{equation}
  \label{Gamma}
  \Gamma = \frac{\mid\vec k\mid\sin\theta}{2\mid\vec q\mid}
          \left(\frac{d\sigma}{d\Omega}\right)^{-1}.
\end{equation}

The isospin decomposition of the physical amplitudes ${\mathcal M}_{l}(N\pi)$
is given by the linear combinations
\begin{eqnarray}
  \label{iso_dec}
  {\mathcal M}_{l}(n\pi^{+}) & = & \sqrt{2}\left({\mathcal M}_{l}^{(0)} +
      {\mathcal M}_{l}^{(-)}\right) = \sqrt{2} \left({\mathcal M}_{l}^{(0)}
      + \frac{1}{3}{\mathcal M}_{l}^{(\frac{1}{2})}
      - \frac{1}{3}{\mathcal M}_{l}^{(\frac{3}{2})}\right), \nonumber \\
  {\mathcal M}_{l}(p\pi^{-}) & = & \sqrt{2}\left({\mathcal M}_{l}^{(0)} -
      {\mathcal M}_{l}^{(-)}\right) = \sqrt{2} \left({\mathcal M}_{l}^{(0)}
      - \frac{1}{3}{\mathcal M}_{l}^{(\frac{1}{2})}
      + \frac{1}{3}{\mathcal M}_{l}^{(\frac{3}{2})}\right), \nonumber \\
  {\mathcal M}_{l}(p\pi^{0}) & = & {\mathcal M}_{l}^{(0)} +
      {\mathcal M}_{l}^{(+)} = {\mathcal M}_{l}^{(0)}
      + \frac{1}{3}{\mathcal M}_{l}^{(\frac{1}{2})}
      + \frac{2}{3}{\mathcal M}_{l}^{(\frac{3}{2})}, \nonumber \\
  {\mathcal M}_{l}(n\pi^{0}) & = & - {\mathcal M}_{l}^{(0)} +
      {\mathcal M}_{l}^{(+)} = - {\mathcal M}_{l}^{(0)}
      + \frac{1}{3}{\mathcal M}_{l}^{(\frac{1}{2})}
      + \frac{2}{3}{\mathcal M}_{l}^{(\frac{3}{2})}. \nonumber \\
\end{eqnarray}

\section{Dispersion relations at fixed \T} \label{sec:disp}

Assuming analyticity and unitarity, the invariant amplitudes $A(s,t)$
for pion photoproduction may be written as the sum of pole contributions
and dispersion integrals. Since the Mandelstam variable
$t$ is kept fixed, the integral runs over the
energy variable $s$ from $\pi N$ threshold to infinity. Though it is generally
difficult to prove the validity of dispersion relations, Oehme and Taylor
\cite{Oeh59} have given such a proof in this case, at least for
sufficiently small values of $t$. As usual in this context, the pole terms
are given by the Born terms in pseudoscalar pion-nucleon coupling. This may 
seem to be at variance with the well-established fact that pseudovector
coupling should be preferred because of chiral symmetry. However, both
approaches lead to the same contributions at the pole, and the violation
of chiral symmetry by using the pseudoscalar coupling for the Born terms will
be removed by appropriate contributions of the dispersion integral.

Starting from these fixed-$t$
dispersion relations for the invariant amplitudes
 of pion photoproduction, the projection of the multipole amplitudes leads
to a well known system of integral equations \cite{Sch69},
\begin{equation} \label{inteq}
\mbox{Re}{\cal M}_{l}(W) = {\cal M}_{l}^{\mbox{\scriptsize P}}(W)
+ \frac{1}{\pi}\sum_{l'}{\cal P}\int_{W_{\mbox{\scriptsize thr}}}^{\infty}
K_{ll'}(W,W')\mbox{Im}{\cal M}_{l'}(W')dW',
\end{equation}
where ${\cal M}_l$ stands for any of the multipoles
$E_{l\pm}, M_{l\pm},$ and ${\cal M}_{l}^{\mbox{\scriptsize P}}$ for the
corresponding (nucleon) pole
term. The kernels $K_{ll'}$ are known, and the real and imaginary parts of the
amplitudes are related by unitarity. In the energy region below two-pion
threshold, unitarity is expressed by the final state theorem of
Watson \cite{Wat54},
\begin{equation} \label{watson}
{\cal M}_{l}^{I} (W) = \mid {\cal M}_{l}^{I} (W)\mid e^{i(\delta_{l}^{I} (W)
+ n\pi)},
\end{equation}
where $\delta_{l}^{I}$ is the corresponding $\pi N$ phase shift and $n$ an
integer. 

For completeness we note that Eq.~(\ref{watson}) assumes that the Compton 
phase may be neglected, which could introduce systematical errors at the
level of one percent. The further numerical solution of Eqs.~(\ref{inteq})
with the constraints imposed by unitarity follows
the method of Schwela et al. 
\cite{Sch69}. In addition
we have taken account of the
coupling to higher states neglected in that earlier reference. At the
energies above two-pion threshold up to $W = 2$ GeV, Eq.~(\ref{watson}) has
been replaced by an ansatz based on unitarity \cite{Sch69}. 
Though this ansatz is  by no means unique, it was motivated by a
comparison with the existing data at the higher energies. Furthermore,
we assume that the contribution of the dispersive integrals from 
$2$ GeV to infinity is largely dominated by $t$-channel exchange. In
earlier references \cite{Kel70} the high-energy behavior has been simulated by
contributions of Regge trajectories. This involved a large number of free
parameters which had to be fitted to the existing data with large error bars.
Since we will restrict our calculations to energies $E_{\gamma} \le$ 450 MeV,
the main $t$-channel contribution should be due to (a fraction of) $\rho$- and
$\omega$-exchange, described by four coupling parameters. 
Furthermore, we have to allow for the addition of solutions of the
homogeneous equations to the coupled system of Eq.~(\ref{inteq}). The whole
procedure introduces
10 free parameters, which have to be determined by a fit to the data. In our
 data base we have included the recent MAMI experiments
for $\pi^{\circ}$ and $\pi^{+}$ production off the proton in the
energy range from 160  MeV to 420 MeV \cite{Fuc96,Kra96,Hae96},
both older and more recent data from Bonn for $\pi^{+}$ production off
the proton \cite{Men77,Bue94,Dut95}, and older Frascati \cite{Car73} and more
recent TRIUMF data
\cite{Bag88} on $\pi^{-}$ production off the neutron. 

In cases of weak coupling between the respective channels 
we neglected some of the integral kernels $K_{ll'}$ in our fitting 
procedure. Therefore we 
iterated the full integral equations (\ref{inteq}) to ensure consistency 
of the method. At low energies the only cases where we found a 
significant discrepancy 
between our original solution and the result after iteration, were 
the $E_{0+}$ 
amplitudes for neutral pion production. For this there are several reasons. 
First of all we expect strong effects from isospin breaking 
and unitarity for the $s$-wave amplitude near threshold. 
In our approach the assumption of isospin 
symmetry is quite essential when we impose the phases of the 
partial waves according to Eq.~(\ref{watson}). However, We introduce 
some isospin 
breaking by hand via the mass difference of charged and neutral pions. 
Second, the $E_{0+}$ amplitudes of neutral pion production are sums of  
large contributions with strong cancelations 
(see Tab.~\ref{tab:e0p}) between the large (pseudoscalar!) Born term
and the dispersive integrals. As a consequence slight variations 
of these contributions can strongly affect the final results.

\section{Results} \label{sec:results}

As can be seen from Tab.~\ref{tab:e0p}, the $s$-waves of charged pion 
production at threshold are mainly given by the pole term contributions, 
while there are only small corrections from 
the dispersion integrals. In particular there is a strong cancelation 
between the (large) $P_{33}$ contributions 
to the $I=\frac{1}{2}$ and $I=\frac{3}{2}$ 
components entering into the linear combination $E_{0+}^{(-)}$ 
(see Eqs.~(\ref{iso_dec})). In 
\cite{Hoe64} this contribution had been determined to be 
0.05$\times 10^{-3}/m_{\pi}$, assuming only the $M_{1+}(\frac{3}{2})$ amplitude 
to contribute. 
According to our result this contribution is 0.09$\times 10^{-3}/m_{\pi}$. More 
interesting is the fact, that the contribution of the $E_{1+}(\frac{3}{2})$
multipole which was neglected in \cite{Hoe64} 
is much larger (0.27$\times 10^{-3}/m_{\pi}$). We also stress that the $\Delta$ 
contribution determined in this way differs even in sign from 
the corresponding value of -0.57$\times10^{-3}/m_{\pi}$ 
which has been derived in 
heavy baryon chiral perturbation theory (HBChPT) \cite{Ber96b}.
However, such differences in individual terms are not unexpected, because
there is no unique correspondence between the graphology of perturbation
and dispersion theory.

\begin{figure}[htbp]
\centerline{\psfig{figure=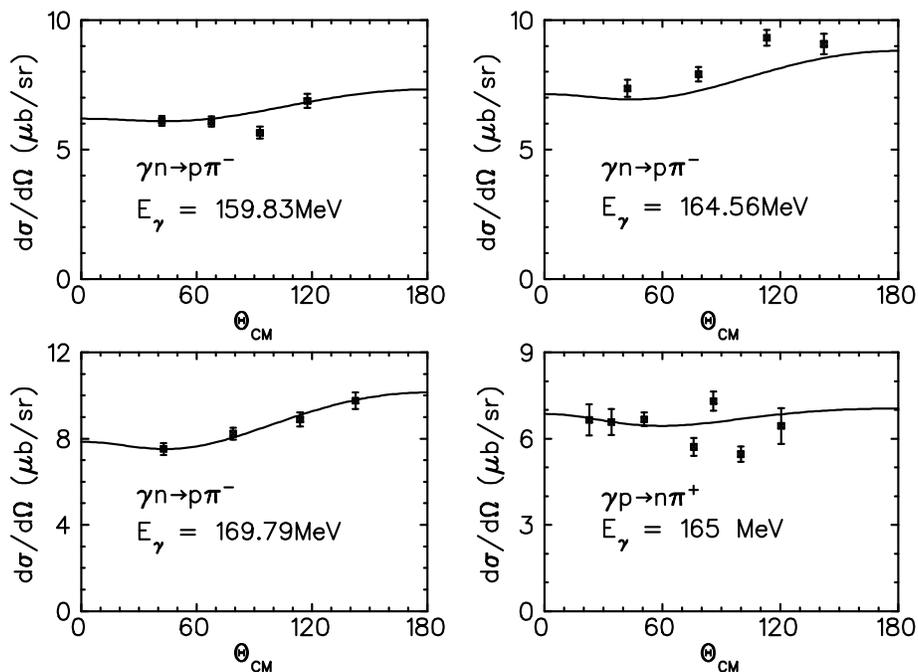,width=12cm}}
    \caption{Differential cross sections for charged pion production. The 
         data for $\gamma n\to \pi^{-}p$ are taken from 
         Ref.~\protect\cite{Kov95}, 
         the data for $\gamma p\to\pi^{+}n$ from 
         Ref.~\protect\cite{Men77}.}
  \label{fig:1}
\end{figure}

The first prediction for the threshold values of charged pion production 
was a LET stating that the $s$-wave multipoles 
are given by the Kroll--Ruderman term \cite{Kro54}, 
$E_{0+}^{\mathrm thr}(\pi^{+}n) = 27.6\times 10^{-3}/m_{\pi^{+}}$, 
$E_{0+}^{\mathrm thr}(\pi^{-}p) = -31.7\times 10^{-3}/m_{\pi^{+}}$, 
where we assumed $g_{\pi N}=13.4$. 
Recently, corrections to these values have been calculated in HBChPT,
to $O(\mu^2)$.  
As can be seen from Tab.~\ref{tab:s_pi_pm} there 
is good agreement for $\pi^{+}$ production between our result and 
ChPT, and both agree nicely with the experiment (see also Fig.~\ref{fig:1}). In 
the case of $\pi^{-}$ production we find agreement of our result with 
the old LET value and the result of an old experimental analysis, while 
ChPT indicates a slightly stronger $E_{0+}$  
multipole (see Tab.~\ref{tab:s_pi_pm}). The preliminary analysis of
the angular distribution in  
a recent experiment from TRIUMF (experiment E643), in which the inverse
reaction 
$\pi^{-}p\to \gamma n$ has been studied, resulted in the much larger value 
$(-34.6\pm 1.0)\times 10^{-3}/m_{\pi^{+}}$ \cite{Kov95}. 
At variance with that finding, the threshold 
extrapolation of the total cross sections determined from the same 
experiment led to 
$(-32.8 \pm 0.7) \times 10^{-3}/m_{\pi^{+}}$, reasonably 
close to the theoretical expectations. 
As has been shown in Fig.~\ref{fig:1}, our analysis describes the 
angular distribution at two energies quite well, 
although our solution for the $E_{0+}$ amplitude is significantly 
smaller than the values according to the analysis of Ref.~\cite{Kov95}. 
We note that these threshold data were not included in our fit, and 
conclude that there must be some inconsistency in the data analysis,
probably due to an underestimate of the statistical errors in the
angular distributions. 
There is special theoretical interest in this amplitude 
because it allows for an independent determination of the difference 
$a_{1}-a_{3}$ of the pion-nucleon scattering lengths via the Panofsky
ratio,
  $P = \sigma(\pi^{-}p\to\pi^{0}n)/\sigma(\pi^{-}p\to\gamma n)$. 
This ratio is well determined by experiment, $P=1.543\pm 0.008$ \cite{Spu77}, 
and related with the scattering lengths by time reversal, 
\begin{equation}
 \label{scat_mul}
  (a_{1}-a_{3})^{2}=9\frac{q}{k_{0}}\mid 
  E_{0+}^{\mathrm thr}(\pi^{-}p)\mid^{2} P,
\end{equation}
where $q = \mid \vec{q} \mid$ and $k_0 = \mid \vec{k_{\pi^0}} \mid$ are the
$cm$ momenta of photon and neutral pion at $p \pi^-$ threshold. 
Using $9q/k_0 = 41.3$, our value of the threshold amplitude, and the measured
Panofsky ratio, we find $a_1 - a_3 = 0.253/m_{\pi}$. This has to be compared 
with the value $0.274 \pm 0.012/m_{\pi}$ resulting from a partial wave 
analysis of pion-nucleon 
scattering \cite{Hoe83} (solution KH80). Recently, $a_1 - a_3$ has also been
determined by studying the level spacing of pionic atoms, with a preliminary
value of $0.288/m_{\pi}$ \cite{Sig95}.

\begin{figure}[htbp]
\centerline{\psfig{figure=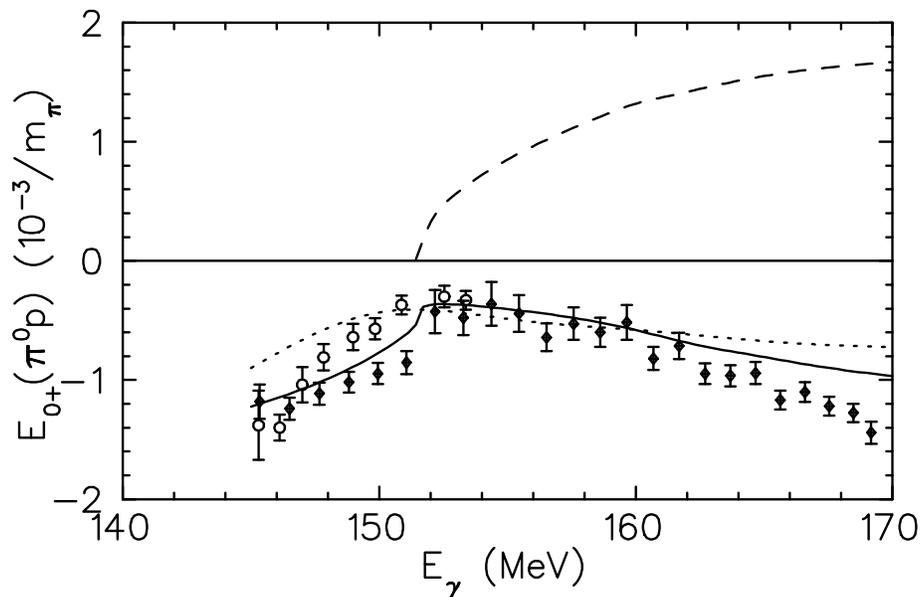,width=12cm}}
    \caption{Real part of the $E_{0+}(\pi^{0}p)$ 
      amplitude before 
      (dotted line) and after (solid line) iteration of 
      Eqs.~(\protect\ref{inteq}). The dashed line is the imaginary part of 
      this amplitude. Data from Ref.~\protect\cite{Fuc96} (circles) 
      and Ref.~\protect\cite{Ber96} (diamonds).}
  \label{fig:2}
\end{figure}

As has been previously mentioned, an exact prediction of the
$s$-wave multipoles for neutral pion production is difficult. The experimental
values show a pronounced cusp effect at the $\pi^+$ threshold, and the expansion
in $\mu$ in HBChPT seems to converge only slowly. In the framework of 
dispersion relations, we observe a delicate cancelation between the large 
negative value of the (pseudoscalar) Born term and the contributions of
the dispersion integrals (see line 3 of Tab.\ref{tab:e0p}). Our final 
threshold value of $-1.22$ (here and in the following 
in units of $10^{-3}/m_{\pi}$) is obtained after
iterating the integral equations, i.e. by inserting the imaginary parts
of our best fit into the integrals on the $rhs$ of Eqs.~(\ref{inteq}). 
As has been mentioned, the solutions 
of the best fit have been obtained by neglecting some of the weak 
couplings in the system of coupled equations. In general, the iterated 
solutions agreed with the original ones at the per cent level, 
thus demonstrating the validity of our approximations. The only exception
to this is the $E_{0+}$ amplitude for neutral pion production, where we 
obtain about $-0.9$ at threshold without the iteration \cite{Han95}. 
As is shown in Fig.~\ref{fig:2}, the iterated solution 
describes the energy dependence of the 
data very well, in particular the steep rise of Im$E_{0+}$ 
at $\pi^{+}$ threshold gives rise to the observed cusp effect. 

In the case of the reaction 
$\gamma n \rightarrow n \pi^0$, we predict a similarly strong cancelation of
Born terms ($-5.2$) and resonance contributions, leading to a final result
of $+ 1.19$ (see Tab.~\ref{tab:e0p}). However, our result 
is considerably lower than the value of $2.1$ predicted by ChPT. 
Tab.~\ref{tab:e0p} also shows the
influence of the Roper multipole $M_{1-}$ on the dispersion integrals.
Its size and even the change of sign for $p \pi^0$ vs. $n \pi^0$ is of great
importance for our final result. As the present data for pion production from
the neutron suffer considerably from systematical and statistical errors,
the predictive power of our calculation is of course much weaker in this
case than for the proton.

\begin{figure}[htbp]
\centerline{\psfig{figure=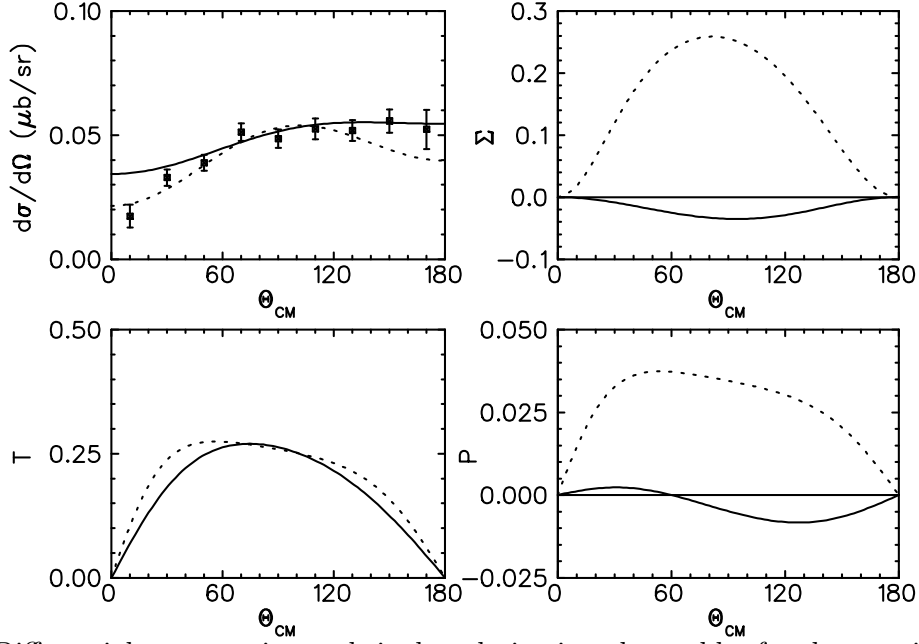,width=12cm}}
    \caption{Differential cross section and single polarization observables 
             for the reaction $\gamma p\to\pi^{0}p$ at 
            $E_{\gamma}=151.69$ MeV (solid lines). The dotted lines show 
            our result after reducing the $M_{1-}$ amplitude by $50 \%$.
            Experimental data from Ref.~\protect\cite{Fuc96}.}
  \label{fig:3}
\end{figure}

\begin{figure}[htbp]
\centerline{\psfig{figure=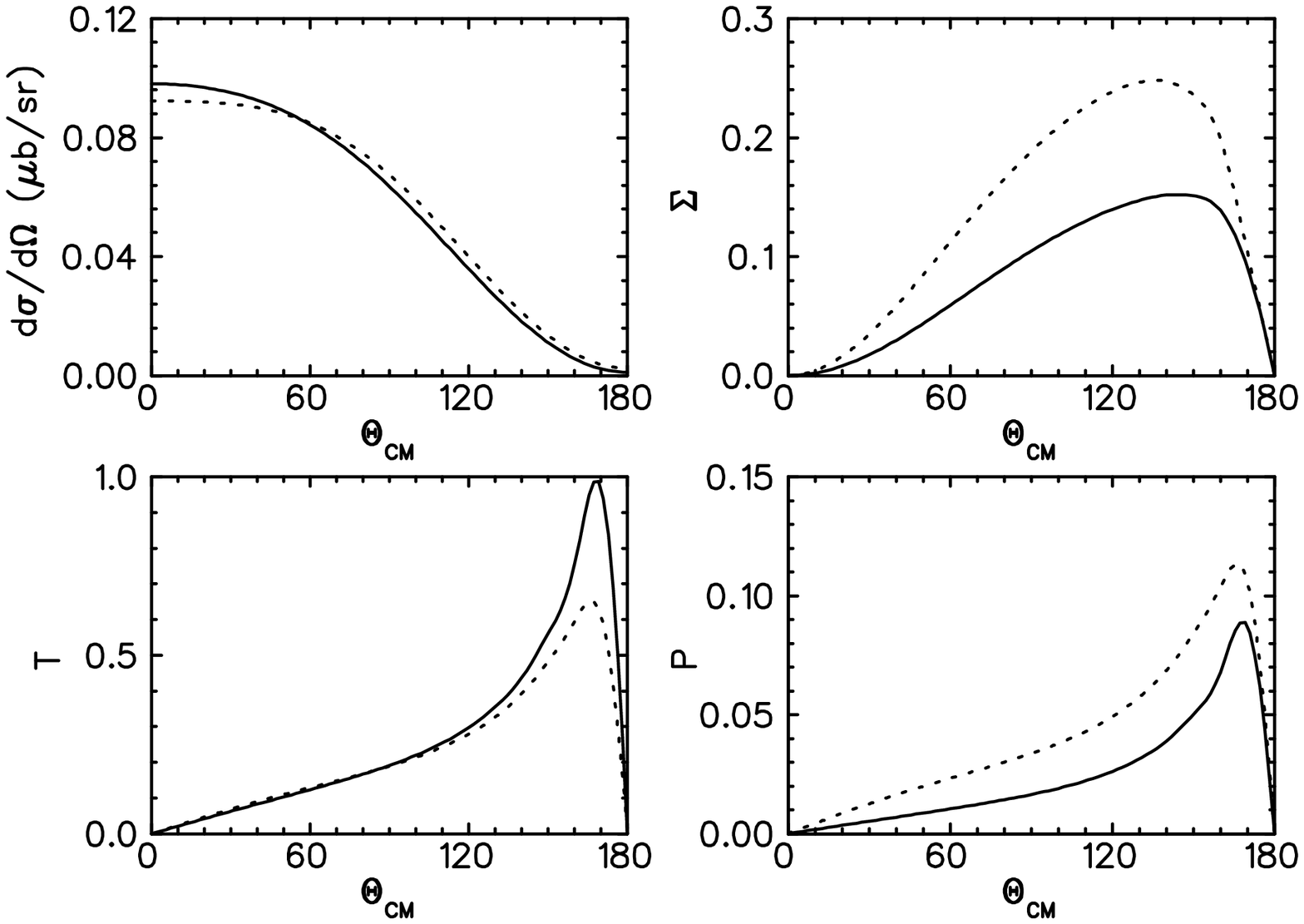,width=12cm}}
    \caption{Differential cross section and single polarization observables 
             for the reaction $\gamma n\to\pi^{0}n$ at 
            $E_{\gamma}=151.69$ MeV (solid lines). The dotted lines show 
            our result after reducing the $M_{1-}$ amplitude by $50 \%$.}
  \label{fig:4}
\end{figure}

In contrast to the $s$-wave amplitudes, the $p$-waves seem to converge much 
faster in HBChPT \cite{Ber96}. 
In particular, it has been possible to derive LETs for the combination
$P_1$ and $P_2$ of Eqs.~(\ref{P1_P3}). Our results for 
these multipoles are compared 
to the predictions of ChPT and experimental analyses in 
Tab.~\ref{tab:s_p_pi0}. There is
general agreement, typically within $5 \%$. The most problematic case is 
again the multipole
$M_{1-}$. As can be seen from Eqs.~(\ref{Sigma}--\ref{rec_pol}), its 
influence is particularly strong
in the case of polarization observables. We have demonstrated this in 
Figs.~\ref{fig:3} and \ref{fig:4} 
by arbitrarily reducing the strength of this multipole by $50 \%$, 
which changes 
even the sign of the beam asymmetry $\Sigma$. 

\section{Conclusion} \label{sec:concl}

In conclusion we find a good overall description of 
the existing threshold data, showing 
the internal consistency of the data at different energies. However, there 
also remain some open questions, particularly regarding the size of the
multipole $M_1$- connected with the Roper resonance, the phases of the 
photoproduction multipoles at energies above the second resonance region, 
and the high-energy behaviour of the amplitudes. 
There is also some doubt concerning the accuracy of the neutron data, and 
it is quite clear 
that the present predictions of dispersion theory cannot be better than the 
data serving as input. The present uncertainties concerning the size of 
the multipole $M_{1-}$ should be removed by the analysis of the recently 
measured beam asymmetry $\Sigma$ for neutral pion production off 
the proton \cite{A2794}. There is also hope that the proposed experiment 
$d(e,e'd)\pi^{0}$ \cite{A1196} will shed some light on the threshold amplitude 
for neutral pion photoproduction off the neutron. This reaction will be an 
extremely sensitive test of model predictions and of the size of isospin 
symmetry breaking in threshold pion production.

\acknowledgments

We would like to thank Prof.~G.~H\"ohler for very fruitful discussions
and the members of the A2 collaboration at Mainz for providing us with
their preliminary data, in particular R.~Beck, F.~H\"arter, H.-P.~Krahn
and H.~Str\"oher.  This work was supported by the Deutsche
Forschungsgemeinschaft (SFB~201).

\begin{table}[htbp]
    \caption{Decomposition of the $E_{0+}$ amplitudes in 
          units of $10^{-3}/m_{\pi^{+}}$. We give the 
           contributions of the pole terms and of the dispersion integrals 
           (\protect\ref{inteq}) over the respective multipoles.}
  \begin{center}
    \leavevmode
    \begin{tabular}{lrrrrrrr}
     & pole & $M_{1+}^{(\frac{3}{2})}$ & $E_{1+}^{(\frac{3}{2})}$
     & $E_{0+}$ & $M_{1-}$ & others & sum \\
    \hline
    $E_{0+}(n\pi^{+})$ & 26.84 & 0.13 & 0.40 & -0.21 & 0.91 & -0.065
                       & 27.99 \\
    $E_{0+}(p\pi^{-})$ & -30.43 & -0.13 & -0.40 & -1.84 & 0.86 & 0.26
                       & -31.67 \\
    $E_{0+}(p\pi^{0})$ & -7.63 & 4.15 & -0.41 & 2.32 & 0.29 & 0.068 
                       & -1.22 \\
    $E_{0+}(n\pi^{0})$ & -5.23 & 4.15 & -0.41 & 3.68 & -0.93 & -0.05 
                       & 1.19 \\
    \end{tabular}
    \label{tab:e0p}
  \end{center}
\end{table}
\begin{table}[htbp]
    \caption{The $E_{0+}$ amplitudes for charged pion photoproduction at 
            threshold in units of $10^{-3}/m_{\pi^{+}}$. 
            Our values are compared with results from chiral
            perturbation theory \protect\cite{Ber96b} and data analysis 
            \protect\cite{Ada70}. We also give our predictions for the reduced 
            $p$-wave multipoles in units of 
            $\mid\vec k\mid\mid\vec q\mid 10^{-3}/m_{\pi^{+}}^{3}$.}
  \begin{center}
    \leavevmode
    \begin{tabular}{lrrrr|rrrr}
    & $\gamma p\to\pi^{+}n$ & & & & $\gamma n\to\pi^{-}p$ & & &\\
  \hline
    & $E_{0+}$ & $m_{1-}$ & $e_{1+}$ &
      $m_{1+}$ & $E_{0+}$ & $m_{1-}$ &
      $e_{1+}$ & $m_{1+}$ \\
  \hline
    disp. & 28.0 & 6.1 & 4.9 & -9.6 & -31.7 & -8.3 & -4.9 & 11.2 \\
    Ref.~\cite{Ber96b} & 28.2$\pm$0.6 & & & & -32.7$\pm$0.6 & & & \\
    Ref.~\cite{Ada70} & 28.3$\pm$0.04 & & & & -31.8$\pm$0.20 & & &
    \end{tabular}
    \label{tab:s_pi_pm}
  \end{center}
\end{table}

\begin{table}[htbp]
    \caption{$s$-- and $p$--wave multipoles of neutral pion photoproduction 
     at threshold in comparison with predictions 
     from ChPT \protect\cite{Ber96a,Ber96c} and results of data 
     analysis \protect\cite{Fuc96,Ber96}. The unit 
     of the $s$--waves is $10^{-3}/m_{\pi^{+}}$, the unit of the reduced 
     $p$--waves is $\mid\vec k\mid\mid\vec q\mid 10^{-3}/m_{\pi^{+}}^{3}$.}
  \begin{center}
    \leavevmode
    \begin{tabular}{lrrrr}
    & disp. & ChPT, Refs.~\cite{Ber96a,Ber96c} & 
     Exp., Ref.~\cite{Fuc96} & Exp., Ref.~\cite{Ber96} \\
   \hline
    $E_{0+}(p\pi^{0})$ & -1.22 & -1.16 & -1.31$\pm$0.08 & -1.32$\pm$0.11 \\
    $m_{1-}(p\pi^{0})$ & -3.91 & -3.21 & & -3.38$\pm$0.26 \\
    $e_{1+}(p\pi^{0})$ & -0.15 & -0.11 & & -0.67$\pm0.15$ \\
    $m_{1+}(p\pi^{0})$ & 7.07 & 7.45 & & 7.44$\pm$0.04 \\
    $(m_{1+} - m_{1-})(p\pi^{0})$ & 10.99 & 10.65 & & 10.82$\pm$0.26 \\
    $p_{1}(p\pi^{0})$ & 10.52 & 10.33 & 10.02$\pm$0.15 & 9.2$\pm$0.3 \\
   \hline
    $E_{0+}(n\pi^{0})$ & 1.19 & 2.13 & \\
    $p_{1}(n\pi^{0})$ & 7.77 & 7.40 & \\
    $p_{2}(n\pi^{0})$ & -8.77 & -8.36 & \\
    \end{tabular}
    \label{tab:s_p_pi0}
  \end{center}
\end{table}



\begin{thebibliography}{99}
\bibitem{Maz86}E.~Mazzucato et al., Phys. Rev. Lett. 57 (1986) 3144.
\bibitem{Bec90}R.~Beck et al., Phys. Rev. Lett. 65 (1990) 1841. 
\bibitem{deB70}P.~de Baenst, Nucl. Phys. B 24 (1970) 633.
\bibitem{Vai72}I.~A.~Vainshtein and V.~I.~Zakharov, Nucl. Phys. B36 (1972) 589.
\bibitem{Ber91}V.~Bernard, J.~Gasser, N.~Kaiser and U.-G.~Mei{\ss}ner, Phys. 
               Lett. B 268 (1991) 219.
\bibitem{Ber96a}V.~Bernard, N.~Kaiser and U.-G.~Mei{\ss}ner, 
                Phys. Lett. B 378 (1996) 337, \\
                N.~Kaiser, Progr.~Part.~Nucl.~Phys. 36 (1996) 119. 
\bibitem{Fuc96}M.~Fuchs et al., Phys. Lett. B 368 (1996) 20, \\
               A.~Bernstein et al., Phys. Rev. C (1997) in print, preprint 
               nucl-ex/9610005.
\bibitem{Ber96}J.~C.~Bergstrom et al., Phys. Rev. C 53 (1996) 1052.
\bibitem{Kro54}N.~M.~Kroll and M.~A.~Ruderman, Phys. Rev. 93 (1954) 233.
\bibitem{Ber96b}V.~Bernard, N.~Kaiser and U.-G.~Mei{\ss}ner, 
                Phys. Lett. B 383 (1996) 116.
\bibitem{Dre92}D.~Drechsel and L.~Tiator, J. Phys. G: Nucl. Part. 
               Phys. 18 (1992) 449.
\bibitem{Oeh59}R.~Oehme and J.~G.~Taylor, Phys. Rev. 113 (1959) 371.
\bibitem{Sch69}D.~Schwela and R.~Weizel, Z. Physik 221 (1969) 71.
\bibitem{Wat54}K.~M.~Watson, Phys. Rev. 95 (1954) 228.
\bibitem{Kel70}B.~H.~Kellett, Nucl. Phys. B 25 (1970) 205.
\bibitem{Kra96}H.-P.~Krahn, PhD. thesis, Mainz (1996).
\bibitem{Hae96}F.~H\"arter, PhD. thesis, Mainz (1996).
\bibitem{Men77} D.~Menze, W.~Pfeil and R.~Wilcke, Compilation of pion
                photoproduction data, Bonn (1977).
\bibitem{Bue94}K.~Buechler et al., Nucl. Phys. A 570 (1994) 580.
\bibitem{Dut95}H.~Dutz, PhD.~thesis, Bonn (1993),\\
               D.~Kr\"amer, PhD.~thesis, Bonn (1993),\\
               B.~Zucht, PhD.~thesis, Bonn (1995).
\bibitem{Car73}F.~Carbonara et al., Nuovo Cim. 13 A (1973) 59.
\bibitem{Bag88}A.~Bagheri et al., Phys. Rev. C 38 (1988) 875.
\bibitem{Hoe64}G.~H\"ohler and W.~Schmidt, Ann. Phys. 28 (1964) 34.
\bibitem{Kov95}Kailin Liu, "Radiative $\pi^{-}p$ capture and the low energy
               theorem", Ph.D. thesis, University of Kentucky (1994).
\bibitem{Spu77}J.~Spuller et al., Phys. Lett. 67 B (1977) 479.
\bibitem{Hoe83}G.~H\"ohler, Pion-Nucleon Scattering, Landoldt-B\"ornstein, 
                 vol. I/9b2, ed. H.~Schopper, Springer (1983).
\bibitem{Sig95}D.~Sigg et al., Phys. Rev. Lett. 75 (1995) 3245.
\bibitem{Han95}O.~Hanstein, L.~Tiator and D.~Drechsel, 
               $\pi N$ Newsletter 10 (1995) 144.
\bibitem{A2794}R.~Beck et al., MAMI Proposal A2/7--94.
\bibitem{A1196}A.~M.~Bernstein, H.~Merkel (spokesmen) et al., 
               MAMI Proposal A1/1--96.
\bibitem{Ber96c}V.~Bernard, N.~Kaiser and U.-G.~Mei{\ss}ner, 
                Z. Phys. C 70 (1996) 483.
\bibitem{Ada70}M.~I.~Adamovich et al., Sov.~J.~Nucl.~Phys. 7 (1970) 360.

\end{thebibliography}
\end{document}